\documentclass[fleqn,usenatbib]{mnras}

\usepackage[T1]{fontenc}
\usepackage{ae,aecompl}

\usepackage{graphicx}	
\usepackage{amsmath}	
\usepackage{amssymb}	
\usepackage{breqn}
\usepackage{color}
\usepackage{cases}
\usepackage{enumerate}
\usepackage{threeparttable}
\usepackage[multidot]{grffile}

\usepackage{txfonts}
\hypersetup{draft}

\usepackage[authoryear]{natbib}
\bibpunct{(}{)}{;}{a}{}{,}
\newcommand\meraxes{{\sc Meraxes}~}
\newcommand\meraxesns{{\sc Meraxes}}
\defcitealias{Qin2017}{Q17}
\defcitealias{Marshall2019}{M19}

\title[DRAGONS XVIII: Black holes \& their host galaxies]{Dark-ages reionization and galaxy formation simulation - XVIII. The high-redshift evolution of black holes and their host galaxies}

\author[M. A. Marshall et al.]{Madeline A. Marshall$^{1,2}$\thanks{E-mail: madelinem1@student.unimelb.edu.au (MAM); swyithe@unimelb.edu.au (JSBW)}, Simon J. Mutch$^{1,2}$, Yuxiang Qin$^{1,3}$, Gregory B. Poole$^4$,
\newauthor J. Stuart B. Wyithe$^{1,2}$\footnotemark[1]
\\
$^{1}$ School of Physics, University of Melbourne, Parkville, VIC 3010, Australia\\
$^{2}$ ARC Centre of Excellence for All Sky Astrophysics in 3 Dimensions (ASTRO 3D)\\
$^{3}$ Scuola Normale Superiore, Piazza dei Cavalieri 7, I-56126 Pisa, PI, Italy\\
$^{4}$ Centre for Astrophysics and Supercomputing, Swinburne University of Technology, PO Box 218, Hawthorn VIC 3122, Australia\\}

\date{Accepted XXX. Received YYY; in original form ZZZ}

\pubyear{2018}

\begin{document}
\label{firstpage}
\pagerange{\pageref{firstpage}--\pageref{lastpage}}
\maketitle

\begin{abstract}
Correlations between black holes and their host galaxies provide insight into what drives black hole--host co-evolution.
We use the \meraxes semi-analytic model to investigate the growth of black holes and their host galaxies from high redshift to the present day. 
Our modelling finds no significant evolution in the black hole--bulge and black hole--total stellar mass relations out to a redshift of 8. 
The black hole--total stellar mass relation has similar but slightly larger scatter than the black hole--bulge relation, with the scatter in both decreasing with increasing redshift. In our modelling the growth of galaxies, bulges and black holes are all tightly related, even at the highest redshifts.
We find that black hole growth is dominated by instability-driven or secular quasar-mode growth and not by merger-driven growth at all redshifts. 
Our model also predicts that disc-dominated galaxies lie on the black hole--total stellar mass relation, but lie offset from the black hole--bulge mass relation, in agreement with recent observations and hydrodynamical simulations.
\end{abstract}




\section{Introduction}
Extensive low-redshift studies reveal a complex interplay between galaxies and the supermassive black holes that reside at their centres, with clear correlations observed between black hole mass and host bulge mass, total stellar mass, velocity dispersion and luminosity (\citealp[e.g.][]{Magorrian1998,Gebhardt2000,Merritt2001,Tremaine2002,Marconi2003,Haring2004,Bentz2009,Kormendy2013,Reines2015}; see the review by \citealp{Heckman2014}).  These tight correlations suggest a co-evolution between galaxies and supermassive black holes, which may be causal, due to feedback from the active galactic nucleus \citep[AGN; e.g.][]{Silk1998,Matteo2005,Bower2006,Ciotti2010} or the efficiency with which the galaxy can fuel the black hole \citep[e.g.][]{Hopkins2010,Cen2015,AnglesAlcazar2017}, or coincidental, simply due to mergers causing both black hole and galaxy growth \citep[e.g.][]{Haehnelt2000,Croton2006b,Peng2007, Gaskell2011,Jahnke2011}. To understand what drives black hole--host co-evolution, it is necessary to study how these correlations change with redshift.

Observing high-redshift black hole--host correlations is fraught with difficulties. Host galaxies are hard to detect since they are often completely outshined by the AGN light, particularly in the rest-frame optical where common stellar mass estimators can be used \citep[e.g.][]{Zibetti2009,Taylor2011}. Subtracting the quasar light has resulted in host detections out to $z\simeq2$ \citep{Jahnke2009,Mechtley2016}, but is yet to be successful for detecting the highest redshift quasars at $z\simeq6$ \citep{Mechtley2012}. For these quasars, host masses are often estimated using the widths of observed submillimeter and millimeter emission lines, such as the [CII]$_{158\mu \textrm{m}}$ and CO (6--5) lines \citep[e.g.][]{Wang2013}.
However, dynamical masses determined from emission line widths are highly dependent on the assumptions made, such as the gas-disc geometries and inclination angles \citep[e.g.][]{Valiante2014}. In fact, inclination angle assumptions can change the determined $M_{\textrm{BH}}/M_{\textrm{bulge}}$ measurements by roughly 3 orders of magnitude \citep{Wang2013}. In addition, the emission regions may not trace the spatial distribution of the stellar component of the galaxy, meaning that these dynamical masses may not be representative of the total stellar mass \citep{Narayanan2009}.
Determining the black hole masses of high-z quasars is also difficult, with emission-line based estimators relying on calibrations at low redshift. Where these observations are unavailable, Eddington accretion rates are instead often assumed to estimate the black hole mass \citep[as in e.g.][]{Wang2013,Willott2017}, which also leads to large uncertainties. High-redshift studies of the black hole--host mass relations are thus very uncertain.

With this in mind, high redshift observations find black holes that are more massive than expected by the local relation, where the canonical black hole--bulge mass ratio is $10^{-2.31\pm0.05}$ for a bulge mass of $10^{11}M_\odot$ \citep{Kormendy2013}. 
For example, ALMA observations of five $z\simeq6$ quasar hosts show black hole to dynamical mass ratios ($M_{\textrm{BH}}/M_{\textrm{dyn}}$) ranging from $10^{-1.9}$ to $10^{-1.5}$ \citep{Wang2013}. Similar studies at $z\simeq4$--7 \citep[e.g.][]{Maiolino2007,Riechers2008,Venemans2012} also give estimates for individual quasars of $M_{\textrm{BH}}/M_{\textrm{dyn}}\gtrsim10^{-2}$, which is significantly larger than the local value if dynamical masses and bulge masses are assumed to be roughly equivalent. This suggests a faster evolution of the first supermassive black holes relative to their host galaxies \citep{Valiante2014}, which could potentially be a result of super-Eddington accretion \citep{Volonteri2015}.

The high observed $M_{\textrm{BH}}/M_{\textrm{dyn}}$ relation at high redshift could, however, be a result of selection effects \citep{Lauer2007,Schulze2011,Schulze2014,DeGraf2015,Willott2017}. \citet{Willott2017} suggest that since only the most massive $z>6$ black holes are observed, if the relation has a wide dispersion then one would expect to see a higher value due to the Lauer bias \citep{Lauer2007}: since the luminosity function falls off rapidly at high masses, the most massive black holes occur more often as outliers in galaxies of smaller masses than as typical black holes in the most massive galaxies.
Indeed, \citet{Willott2017} found that $M_{\textrm{BH}}<10^9M_\odot$ black holes at $z>6$ fall below the $M_{\textrm{BH}}$--$M_{\textrm{dyn}}$ relation for low redshift galaxies, in contrast to the opposite being true for higher mass black holes. 
Similarly, \citet{Schulze2014} claim that selection effects are the reason for the observed evolution of the $M_{\textrm{BH}}$--$M_{\textrm{Bulge}}$ relation; on applying a fitting method to correct for selection effects, they find no statistical evidence for a cosmological evolution in the $M_{\textrm{BH}}$--$M_{\textrm{Bulge}}$ relation.

A lack of evolution in the black hole--host relations is consistent with the findings of cosmological hydrodynamical simulations such as Horizon-AGN \citep{Volonteri2016}, which observes very little evolution in the $M_{\textrm{BH}}$--$M_\ast$ relation from $z=0$ to 5, and {\sc BlueTides} \citep{Huang2018}, which finds a $M_{\textrm{BH}}$--$M_\ast$ relation at $z=8$ that is consistent with the local \citet{Kormendy2013} relation. \citet{DeGraf2015}, on the other hand, found that the relation evolves slightly for $z\geq1$ for the highest mass black holes, with a steeper slope at the high-mass end at higher redshifts, making selection effects important. The more statistical study of \citet{Schindler2016} found that the ratio of the black hole to stellar mass density is constant within the uncertainties from $z=0$ to 5, with a slight decrease in the ratio at $3\leq z\leq5$; this is also consistent with no cosmological evolution in the $M_{\textrm{BH}}$--$M_\ast$ relation.

In this work we explore the evolution of the black hole--host relations with the \meraxes semi-analytic model \citep{Mutch2016}. 
\meraxes is designed specifically to study galaxy formation and evolution at high redshifts, making it ideal for studying the evolution of black holes and their host galaxies. 
The outline of the paper is as follows. We give a brief overview of \meraxes in Section \ref{Meraxes}, and detail the calibration procedure in Section \ref{Verification}. We then investigate the evolution of black holes in the model in Section \ref{Results}, and conclude in Section \ref{Conclusion}.
Throughout this work, we adopt the \citet{Planck2015} cosmological parameters: ($h, ~\Omega_m, ~\Omega_b, ~\Omega_\Lambda, \sigma_8, n_s$)=(0.678, 0.308, 0.0484, 0.692, 0.815, 0.968).

\section{Semi-analytic model}
\label{Meraxes}
In this work we use \meraxesns, a semi-analytic model designed to study galaxy evolution at high redshifts  \citep{Mutch2016}. Using the properties of dark matter halos from an N-body simulation, \meraxes analytically models the physics involved in galaxy formation and evolution.

\subsection{N-body simulations}
We run \meraxes on the collisionless N-body simulations \textit{Tiamat} and \textit{Tiamat-125-HR} \citep{Poole2016,Poole2017}. 
\textit{Tiamat} is ideal for studying high redshifts, with a high mass and temporal resolution.
\textit{Tiamat} runs from $z=35$ to $z=1.8$, with a box size of ($67.8 h^{-1}$ Mpc)$^3$, $2160^3$ particles of mass $2.64\times10^6 h^{-1} M_\odot$, and a high cadence of 11.1 Myr per output snapshot at $z>5$.
\textit{Tiamat-125-HR} is a low-redshift counterpart to \textit{Tiamat}, running from $z=35$ to $z=0$ with the same temporal resolution, but with a lower mass resolution ($1080^3$ particles of mass $1.33\times10^8 h^{-1} M_\odot$) and larger box size of $(125 h^{-1}$ Mpc)$^3$, more suited for low-redshift studies. For a detailed description of these simulations, see \citet{Poole2016} and \citet{Poole2017}. 
Throughout this work, we use the higher resolution \textit{Tiamat} at high-redshifts, and \textit{Tiamat-125-HR} for $z<2$, unless otherwise specified. 

\subsection{The {\sc Meraxes} model}
\meraxes assumes that galaxies reside in the centre of dark matter haloes produced by the N-body simulation. Using the properties of these haloes, \meraxes analytically models the baryonic physics involved in galaxy formation and evolution, such as gas cooling, star formation, black hole growth, and supernova and black hole feedback. These analytical prescriptions involve a range of free parameters, which must be calibrated using observations such as the stellar mass function (see Section \ref{Verification} for details).
The model outputs a range of properties for each galaxy in the simulation, including the mass of hot gas, cold gas and stars, its star formation rate, and the mass of its central black hole. For a full description of the processes modelled in \meraxesns, see \citet{Mutch2016}, \citet{Qin2017} (herein \citetalias{Qin2017}) and \citet{Marshall2019} (herein \citetalias{Marshall2019}). We outline the physical processes most relevant to this work in Sections \ref{sec:BulgeGrowth} and \ref{sec:BHGrowth} below.

\subsection{Bulge growth}
\label{sec:BulgeGrowth}
In \meraxesns, stars in galaxies reside in three components: an exponential disc, a spheroidal merger-driven bulge and a disc-like instability-driven bulge. Bulges grow through both galaxy-galaxy mergers and disc-instabilities. A full description of this model is given in \citetalias{Marshall2019}, with a brief summary outlined below.

\textbf{\textit{Galaxy mergers:} }In \meraxesns, we assume that galaxy mergers with merger ratio $\gamma=M_{\textrm{primary}}/M_{\textrm{secondary}}>0.01$ trigger a burst of star formation, by causing shocks and turbulence in the cold gas of the parent galaxy. The galaxy will also accumulate the mass of the secondary galaxy. We assume that the dominant mass component of the primary galaxy will regulate where these stars produced by the burst and the secondary's mass will be deposited. If the primary is dominated by a discy component (either the stellar disc or instability-driven bulge), the mass will be deposited in the plane of the disc and so it is added to the instability-driven bulge. Otherwise, we assume that the new stars will accumulate in shells around the spheroidal merger-driven bulge, and so their mass is added there.

In major mergers, where $\gamma>0.1$ or $\gamma>0.3$ (see Section \ref{Verification}), we assume that the stellar disc and instability-driven bulges are destroyed, with all stars placed into the merger-driven bulge.

\textbf{\textit{Disc instabilities:}} In our model we assume that the galaxy discs are thin, with an exponential surface density and flat rotation curve. Such discs become unstable if $M_{\textrm{disc}}>V_{\textrm{disc}}^2 R_s/G =M_{\textrm{crit}}$ \citep{Efstathiou1982,Mo1998}. Here, we take $M_{\textrm{disc}}$ as the combined mass of both gas and stars in the disc, and $V_{\textrm{disc}}$ and $R_s$ as the mass-weighted velocity and scale radius of the stellar and gas discs. If such a disc instability occurs, \meraxes returns the disc to stability by transferring $M_{\textrm{unstable}}=M_{\textrm{disc}}-M_{\textrm{crit}}$ of stars from the disc to the instability-driven bulge.

\subsection{Black hole growth}
\label{sec:BHGrowth}
The \meraxes black hole model was introduced in \citetalias{Qin2017}, and updated to include instability-driven growth in \citetalias{Marshall2019}. We summarize the model below, however the interested reader is encouraged to refer to \citetalias{Qin2017} for the full details.

In \meraxesns, black holes are seeded in every newly-formed galaxy, with a seed mass of $10^4M_\odot$. Black holes then grow by accretion of both hot and cold gas, through the radio- and quasar modes, respectively. We also assume that black holes grow in galaxy mergers, with the black holes in each galaxy merging together.

\textbf{\textit{Radio mode:}} Black holes accrete hot gas from the static hot gas reservoir around the galaxy (of mass $M_{\textrm{hot}}$ and density $\rho_{\textrm{hot}}$), at a fraction $k_h$ of the Bondi-Hoyle accretion rate:
\begin{equation}
\dot{M}_{\textrm{Bondi}} =\frac{2.5\pi G^2}{c_s^3}  M_{\textrm{BH}}^2 \rho_{\textrm{hot}}.
\end{equation}
We consider $k_h$ a free parameter, which adjusts the efficiency of radio-mode black hole growth \citep{Croton2016}.  This accretion is limited by the amount of hot gas in the reservoir and the Eddington limit, so the mass available for accretion during a simulation time-step of width $\Delta t$ is
\begin{equation}
{M}_{\textrm{accretion}}=\min \left(M_{\textrm{hot}},M_{\textrm{Edd}}, k_h \dot{M}_{\textrm{Bondi}} \Delta t \right),
\end{equation}
where $M_{\textrm{Edd}}$ is the mass that would be accreted continually at the Eddington rate over $\Delta t$:
\begin{equation}
M_{\textrm{Edd}}=M_{\textrm{BH}}\left[\exp \left( \frac{\Delta t}{\eta t_{\textrm{Edd}}}\right) -1 \right].
\label{eq:Edd}
\end{equation}
Here $t_{\textrm{Edd}}\equiv\frac{\sigma_T c}{4 \pi G m_p}\simeq 450$ Myr is the Eddington accretion time,   $M_{\textrm{BH}}$ is the black hole mass at the beginning of the time step, and $\sigma_T$ is the Thomson cross-section.
A fraction $\eta$ of this accretion mass is radiated away---
$L_{\textrm{AGN}}=\eta {M}_{\textrm{accretion}}  c^2$---and so during one snapshot, black holes grow through the radio-mode by mass
\begin{equation}
\Delta {M}_{\textrm{BH,R}} = (1-\eta) {M}_{\textrm{accretion}}.
\label{eq:radio}
\end{equation}

We include the effects of radio-mode AGN feedback by assuming that a fraction $\kappa_r$ of the radiated energy is coupled to the surrounding gas, adiabatically heating a mass of
\begin{equation}
M_{\textrm{heat}}=\frac{\kappa_r  L_{\textrm{AGN}}}{0.5V_{\textrm{vir}}^2}=\frac{\kappa_r  \eta {M}_{\textrm{accretion}}  c^2}{0.5V_{\textrm{vir}}^2}.
\end{equation}
This heated gas is subtracted from the cooling flow, regulating the accretion of new gas onto the black hole \citep[see][Q17]{Croton2006a,Croton2016}.
This AGN feedback has no significant effect on the results of \textit{Tiamat} at $z\geq2$, suppressing the growth of only the most massive galaxies in \textit{Tiamat-125-HR} at lower redshifts  (see Appendix \ref{sec:Appendix}).

\textbf{\textit{Quasar mode:}} Black holes accrete cold gas from the galaxy (total mass $M_{\textrm{cold}}$), when triggered by either a galaxy-galaxy merger or a disc instability. During such an event, the black hole mass grows by a total of
\begin{equation}
\Delta {M}_{\textrm{BH,Q}} =\min \left( M_{\textrm{cold}}, \frac{k M_{\textrm{cold}}}{\left(1+\frac{280\textrm{ km s}^{-1}}{V_{\textrm{vir}}}\right)^2} \right)
\label{eq:BHgrowth}
\end{equation}
where $V_{\textrm{vir}}$ is the virial velocity of the halo and $k$ is a free parameter to adjust the growth efficiency. For merger-triggered growth, we take $k=k_c \gamma$ where $\gamma$ is the merger ratio and $k_c$ is a constant. For instability driven growth, $k=k_i$. We consider $k_c$ and $k_i$ two separate free parameters  \citepalias[see][and Section \ref{Verification}]{Marshall2019}.  During the quasar mode, black holes are assumed to accrete at the Eddington rate, and thus the mass accreted by the black hole during one simulation snapshot is limited to $M_{\textrm{Edd}}$. This can result in the mass $\Delta {M}_{\textrm{BH,Q}}$ being accreted over multiple simulation snapshots (see \citetalias{Qin2017} for a more detailed discussion).

We incorporate quasar-mode AGN feedback by considering the energy injected into the gas during a simulation time-step, $\kappa_q  \eta \min{\left(M_{\textrm{Edd}},\Delta {M}_{\textrm{BH,Q}}\right)}  c^2$,
where $\kappa_q$ is the mass coupling factor.
We assume that this energy generates a wind that heats the cold disc gas and transfers it to the hot gas reservoir, depleting the supply of cold gas available for the black hole to accrete. If sufficient energy is injected by the quasar, this wind can also eject the hot gas (see \citetalias{Qin2017}).

\subsection{Quasar luminosity functions}
\label{sec:QLF_calc}
We calculate the bolometric luminosities of each black hole in the model following the \citetalias{Qin2017} method, which assumes Eddington luminosity for all accreting black holes, and self-consistently calculates the duty cycle. We consider the luminosities from both the quasar- and radio-modes of accretion. As described in \citetalias{Qin2017}, at high-redshifts the contribution from the radio-mode is negligible. At the lowest redshifts ($z\leq2$), the radio-mode becomes a more significant growth mechanism for the most massive black holes, and so their luminosities are enhanced slightly by the addition of the radio-mode luminosity.

We convert from bolometric to $B$-band luminosities using the \citet{Hopkins2007} bolometric correction, and then assume a continuum slope of $\alpha=0.44$ to convert to UV luminosities (see \citetalias{Qin2017} for details). We also account for obscuration due to quasar orientation, by scaling the UV luminosity function by $1-\cos (\theta/2)$, where $\theta$ represents the opening angle of quasar radiation. In our model we assume a constant $\theta$, for simplicity, which is a free parameter in our model; this simply adjusts the normalisation of our UV luminosity functions.

\section{Model calibration}
\label{Verification}
\begin{figure*}
\begin{center}
\includegraphics[scale=0.9]{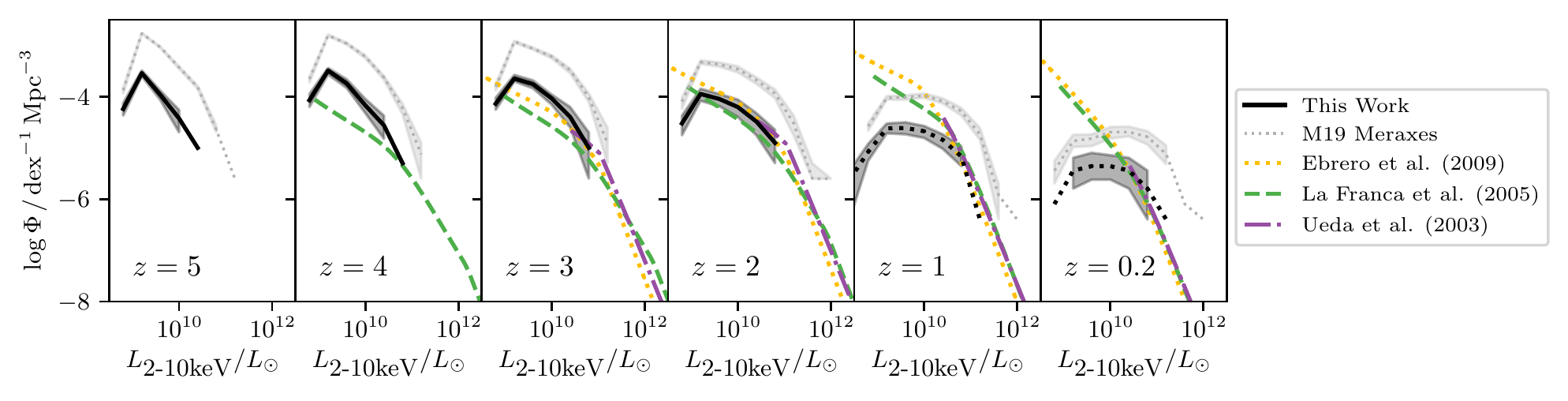}
\caption{Quasar X-ray luminosity functions at $z=5$ to 0.2 from our best \meraxes model applied to \textit{Tiamat} (black solid) and \textit{Tiamat-125-HR} (black dotted). A range of observations are also shown \citep[][see legend]{Ueda2003,Franca2005,Ebrero2009} at redshifts where data was taken; the curves are best-fitting relations using a luminosity-dependent density evolution (LDDE) model. We also plot the \citetalias{Marshall2019} \meraxes model (grey dotted), showing that this overpredicted the observed quasar luminosity functions.}
\label{XrayQLF}
\end{center}
\end{figure*}

\begin{figure*}
\begin{center}
\includegraphics[scale=0.9]{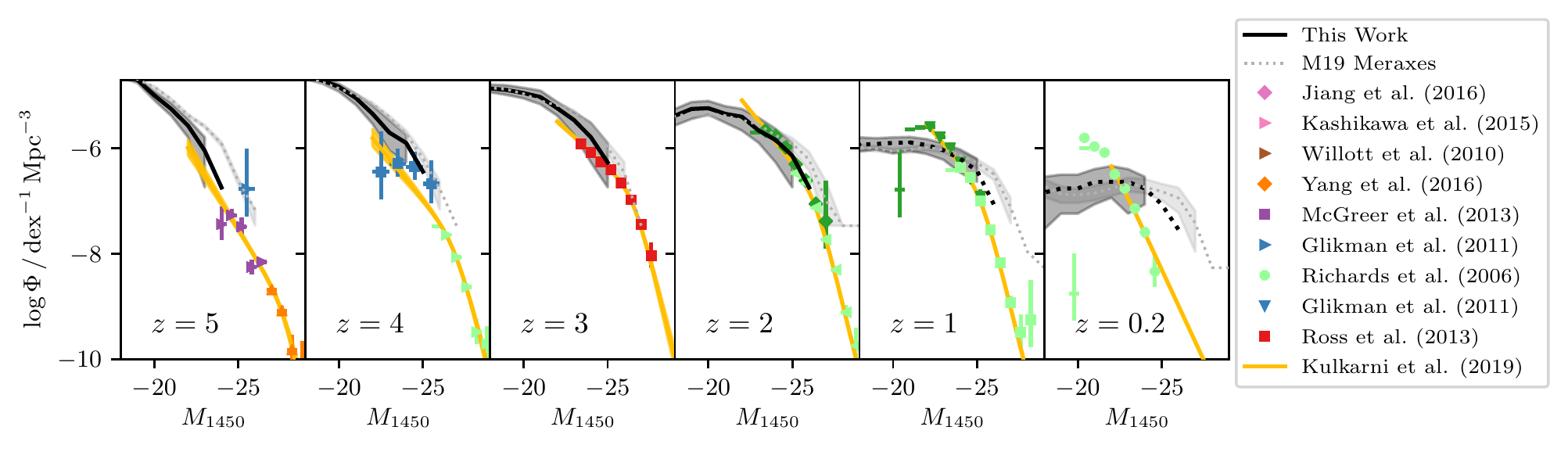}
\caption{Quasar UV luminosity functions at $z=5$ to 0.2 from our best \meraxes model applied to \textit{Tiamat} (black solid) and \textit{Tiamat-125-HR} (black dotted), alongside a range of observations \citep[][see legend]{Richards2006,Willott2010a,Glikman2011,Ross2013,McGreer2013,Kashikawa2015,Yang2016,Jiang2016,Kulkarni2018}. 
We also plot the \citetalias{Marshall2019} \meraxes model (grey dotted), showing that this overpredicted the observed quasar luminosity functions.}
\label{UVQLF}
\end{center}
\end{figure*}

In \citetalias{Marshall2019} we calibrated the free parameters in \meraxes to match the observed stellar mass functions at $z=0$--8 \citepalias[][figure 1]{Marshall2019}, and the black hole--bulge mass relation at $z=0$ \citepalias[][figure 2]{Marshall2019}. Using this model, we find that the black hole mass function and quasar luminosity functions are much larger than predicted by the observations (Figures \ref{BHMF}, \ref{XrayQLF} and  \ref{UVQLF}). 
In addition, we note that \citet{Shankar2016} find significant selection biases in the black hole--bulge mass relation---a topic of recent debate \citep[see e.g.][]{Kormendy2019}. Due to the \citetalias{Marshall2019} predictions and this potential bias, we assume that the \citet{Shankar2009} $z=0$ black hole mass function is a less biased indicator of the local black hole population, and retune the model here to better reproduce the black hole observations. 

Note that we use the same parameter values for \textit{Tiamat} and \textit{Tiamat-125-HR}, and use both simulations to tune the model: \textit{Tiamat} for matching $z\geq2$ observations and \textit{Tiamat-125-HR} for $z<2$. We find that our results from the two simulations are generally consistent at $z\simeq2$, with broad qualitative agreement at higher redshifts, and so we can reliably use the \textit{Tiamat-125-HR} simulation at $z<2$ where \textit{Tiamat} is unavailable (see Appendix \ref{sec:Appendix2} for further discussion).

\begin{figure*}
\begin{center}
\includegraphics[scale=0.9]{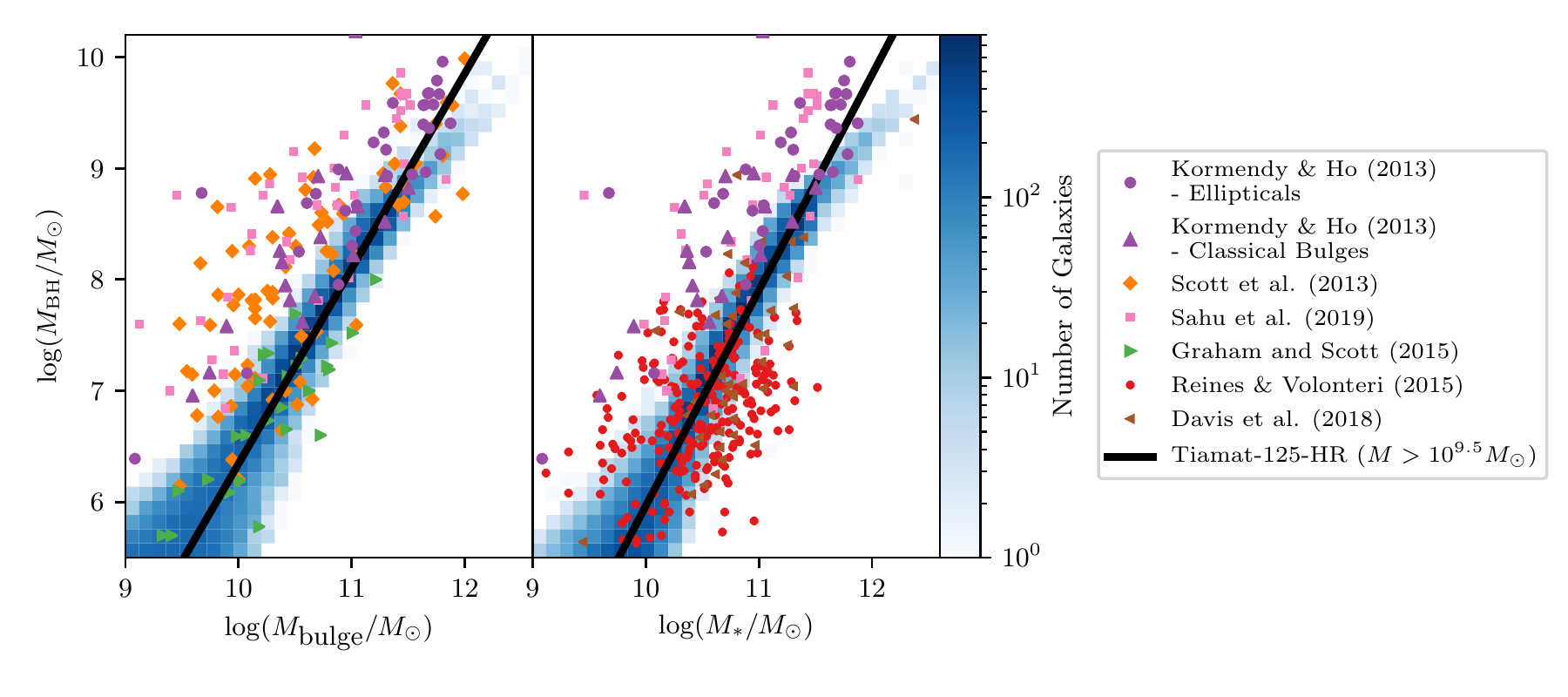}
\caption{\textit{Left panel:} The $z=0$ black hole--bulge mass relation, and \textit{Right panel:} the $z=0$ black hole--total stellar mass relation, for our model applied to \textit{Tiamat-125-HR} (blue density plot). Only galaxies classified as centrals are shown.
A range of observations are also plotted \citep[][see legend]{Kormendy2013,Scott2013,Graham2015,Reines2015,Davis2018,Sahu2019}. A best-fitting line for our model galaxies with $M>10^{9.5} M_\odot$ is also shown (solid line).}
\label{MBHMStellar}
\end{center}
\end{figure*}

We calibrate the free parameters in the model to match the observed stellar mass functions at $z=0$--8 (Figure \ref{SMF}), the \citet{Shankar2009}  and \citet{Davis2014} black hole mass function at $z=0$ (Figure \ref{BHMF}; see Appendix \ref{sec:Appendix}), and the quasar X-ray luminosity functions from $z=5$ to 2 (Figure \ref{XrayQLF}).  Since  \citet{Shankar2016} find that the observed black hole--bulge mass relation is biased to high black hole masses, we also require our model to not over-predict this relation, however we do not otherwise tune to it. We note that our best models produce black hole--host mass relations lower than the observations, consistent with the expectations of \citet{Shankar2009}, and have steeper slopes (Figure \ref{MBHMStellar}).
We find that these criteria are met by a range of free parameter values for the merger-driven black hole growth efficiency, $k_c = 0.005$, 0.01, 0.03 and 0.09, and the definition of a major merger, $\gamma>0.1$ and $\gamma>0.3$. 
We note that all of these parameter sets produce very similar results, so unless otherwise specified we only show the model results for the $k_c=0.005$ and $\gamma>0.1$ case hereafter. 
 
As a further check of the black hole population, we plot the black hole accretion rate density as a function of redshift for models with these different merger-driven black hole growth efficiencies (with $\gamma>0.1$), in Figure \ref{BHARD}. We find that the models with $k_c=0.005$ and $k_c=0.01$ give black hole accretion histories in approximate agreement with the observations (Figure \ref{BHARD}).
The larger values of $k_c$ overproduce measurements of the black hole accretion rate density \citep[e.g.][]{Delvecchio2014}.

\begin{table*}
\centering
\caption{\meraxes black hole growth parameters as used in \citetalias{Marshall2019}, and as retuned for this study.}
\label{tab:Parameters}
\begin{threeparttable}
\begin{tabular}{ccc}
\hline 
Parameter & \citetalias{Marshall2019} & This work \\ 
\hline 
Minimum merger ratio for major merger & 0.1 & 0.1, 0.3 \\ 
Black hole seed mass ($M_\odot$) & $10^4$ & $10^4$ \\ 
Merger-driven black hole growth efficiency\tnote{a}, $~k_c$ & 0.03 & 0.005, 0.01, 0.03, 0.09 \\ 
Instability-driven black hole growth efficiency\tnote{a}, $~k_i$  & 0.02 & 0.005\\ 
Radio mode black hole growth efficiency\tnote{b}, $~k_h$   & 0.003  & 0.03 \\ 
Black hole efficiency of converting mass to energy\tnote{c}, $~\eta$  & 0.2  & 0.06 \\ 
Opening angle of AGN radiation\tnote{d}, $~\theta$  & $30^{\circ} $ & $70^{\circ} $ \\ 
\hline 
\end{tabular} 
\begin{tablenotes}
\item[a] Equation \ref{eq:BHgrowth}
\item[b] Equation \ref{eq:radio}
\item[c] $L_{\textrm{AGN}}=\eta \Delta {M}_{\textrm{BH}}  c^2$, Equation \ref{eq:radio}
\item[d] Section \ref{sec:QLF_calc}
\end{tablenotes}
\end{threeparttable}
\end{table*}

\subsection{Quasar luminosity functions}
\label{sec:QLFs}
The opening angle of AGN radiation, $\theta$, adjusts the normalization of the UV luminosity function. We tune this to match the observations, shown in Figure \ref{UVQLF}, finding a preferred $\theta$ of 70 degrees, corresponding to an observable fraction of UV quasars of 18 per cent.

We show the quasar X-ray luminosity functions at $z=5$--0 in Figure \ref{XrayQLF}, with X-ray luminosities calculated using the \citet{Hopkins2007} bolometric to X-ray correction. At $z=2$ the model and the observations agree remarkably well. At $z>2$ the model over-predicts the observed quasar X-ray luminosity function at intermediate luminosities, by up to $\sim0.7$ dex at $z=4$, while at $z<2$ the model under-predicts the luminosity function at these luminosities. Our model shows better agreement with the observations than previous versions of \meraxes (\citetalias{Marshall2019}, as seen in Figure \ref{XrayQLF}, and \citetalias{Qin2017}; see also \citealt{Amarantidis2019}). 

While the observations show a slight increase in the X-ray quasar luminosity functions from $z=4$ to 2, the model predicts a slight decrease. In fact, we cannot find a combination of black hole parameters (see Table \ref{tab:Parameters}) that results in a redshift evolution that matches that of the observed X-ray quasar luminosity function at $z>2$. However, the key quantity of black hole accretion rate density is predicted by the model to peak at $z=2$ as observed. In addition to published uncertainties in the observations, it may also be the case that at higher redshifts X-ray AGN are more likely to be obscured, which is consistent with evidence from a range of X-ray observations \citep{Treister2006,Vito2014,Buchner2015}. Thus we argue that the inability of our model to match the redshift evolution of the X-ray quasar luminosity function may not represent a significant concern.

\begin{figure}
\begin{center}
\includegraphics[scale=0.9]{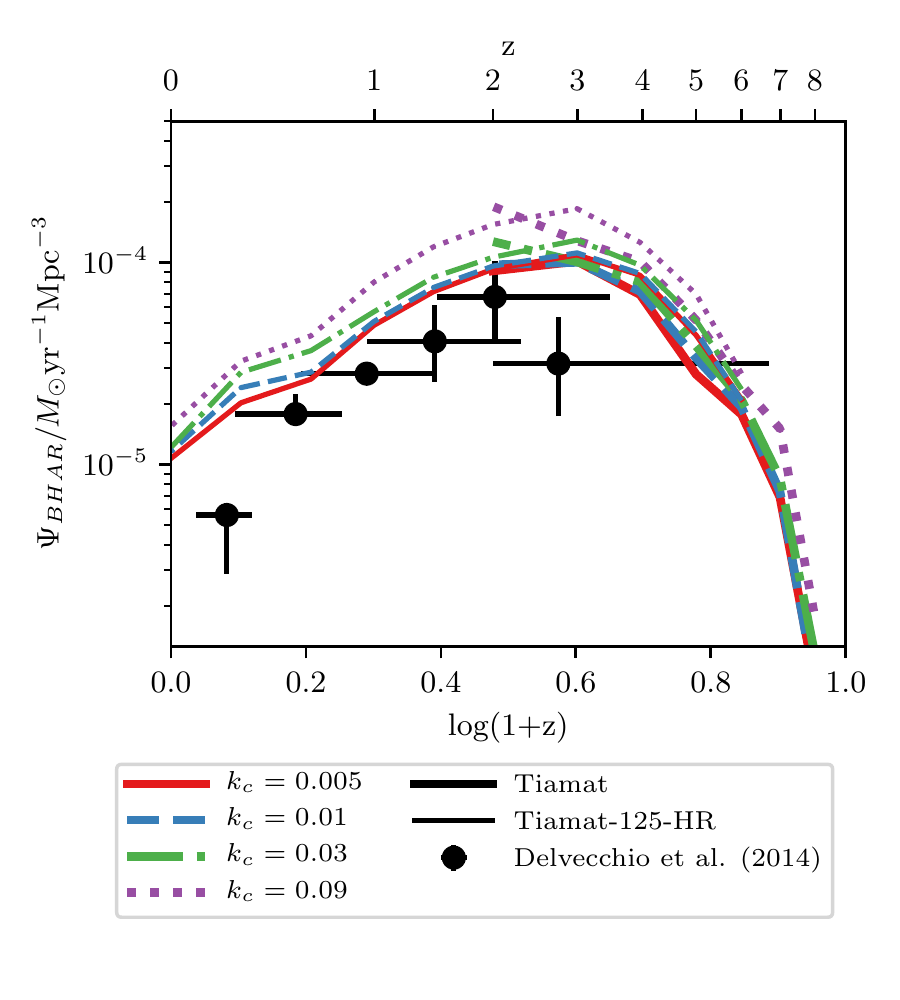}
\caption{The black hole accretion rate density as a function of redshift from \meraxesns, and as estimated from the AGN bolometric luminosity function \citep[black points;][]{Delvecchio2014}. We calculate the black hole accretion rate density as the total black hole mass growth in the simulation between adjacent simulation snapshots, divided by the time between snapshots and normalized by the simulation volume. We show four models, with different merger-driven black hole growth efficiencies: $k_c=0.005$, 0.01, 0.03 and 0.09 (see legend). These parameters were all found during the model tuning to reproduce the observations well.}
\label{BHARD}
\end{center}
\end{figure}

We show the quasar UV luminosity functions at $z=5$--0 in Figure \ref{UVQLF}. We find that, as with the X-ray luminosity function, the UV luminosity function decreases from $z=5$ to 0, though it agrees well with observations at $z>2$. At $z<2$, however, we note that the faint-end of the UV luminosity function becomes flat, and by $z<1$ there is a significant disagreement with the observations, with the model producing too many luminous quasars.  As seen in Figure \ref{BHARD}, the black hole accretion rate density becomes significantly higher than the observations at $z<1$, consistent with the quasar luminosities being overestimated at these redshifts.
This excess black hole accretion is most likely a result of the model missing important physics required for modelling low-redshift galaxy evolution, particularly in the quenching of massive galaxies, or due to the simplifications assumed in the model such as a constant black hole accretion efficiency.
However, as the overall accretion rate density at these redshifts is low, this will not have a significant impact on the black hole mass, an integrated quantity.
Thus, while the $z<1$ black hole accretion rates are overestimated, the black hole mass function (Figure \ref{BHMF}) and black hole--host mass relations (Figure \ref{MBHMStellar}) are reliable at low redshifts.



\section{Results}
\label{Results}
We now use the model described in Sections \ref{Meraxes} and \ref{Verification} to explore black hole growth. We investigate the redshift evolution of the black hole--host scaling relations in Section \ref{sec:RedshiftEvolution}. In Section \ref{sec:BHGrowthModes} we consider the relative contributions of the different black hole growth modes, and in Section \ref{sec:Morphology} we consider the black hole--host scaling relations in galaxies of different morphologies.

\subsection{Redshift evolution of the black hole--bulge and total stellar mass relations}
\label{sec:RedshiftEvolution}
\begin{figure*}
\begin{center}
\includegraphics[scale=0.9]{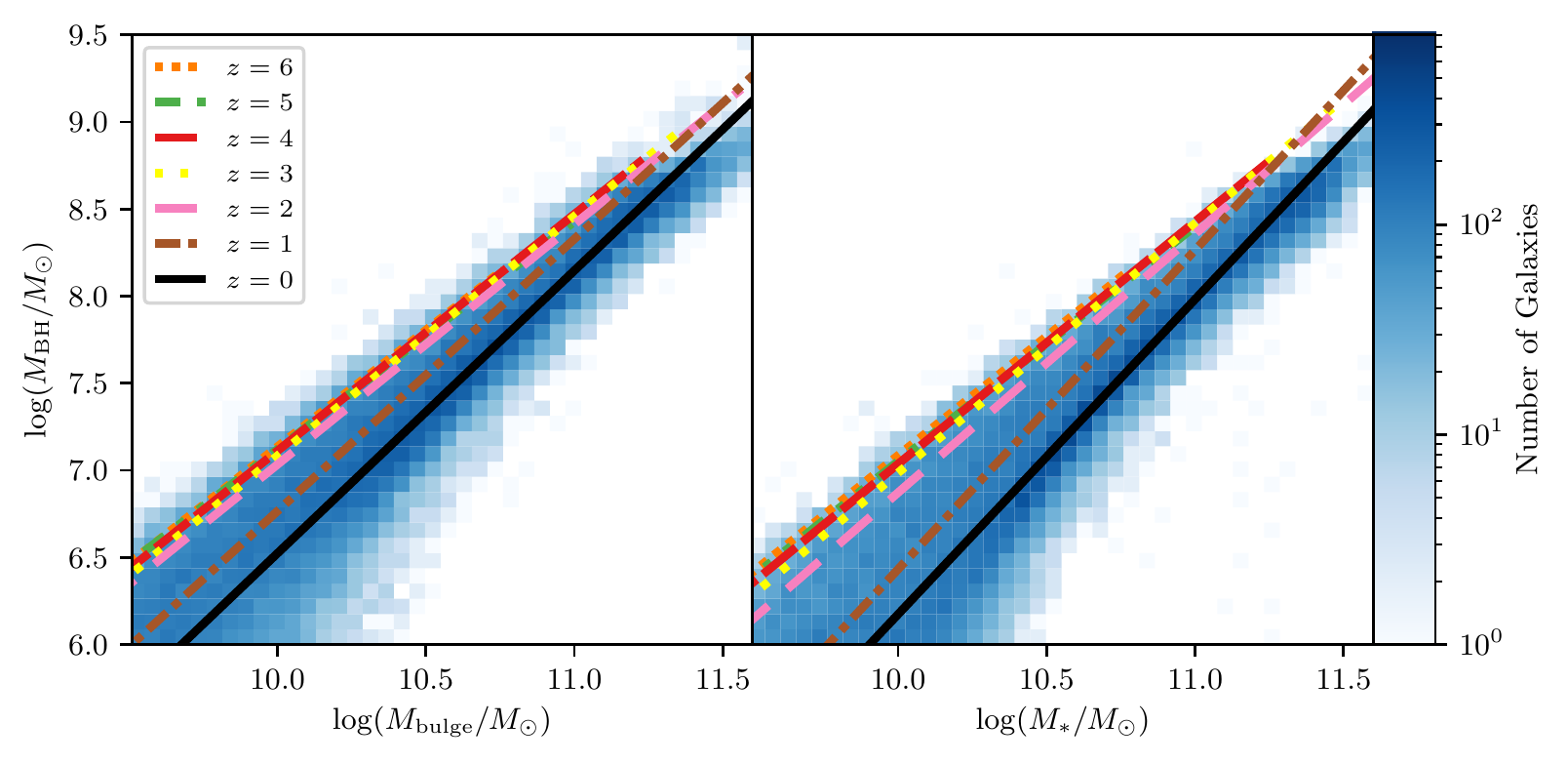}
\caption{Lines of best-fit to the black hole--bulge mass \textit{(left panel)} and black hole--total stellar mass \textit{(right panel)} relations at a range of redshifts, as given by Equation \ref{eq:fits} and the parameters in Table \ref{tab:fits}. The blue density plot shows the $z=0$ distribution. The slope and normalization of these relations increase slightly from $z=0$ to 2, with slow evolution at $z>2$. This evolution is mild relative to the scatter in the relation.
}
\label{Magorrian_evolution}
\end{center}
\end{figure*}

To investigate the redshift evolution of the black hole--bulge and black hole--total stellar mass relations we first perform linear least squares fits to the relations:
\begin{equation}
\log{\left(\frac{M_{\textrm{BH}}}{M_\odot}\right)} = \alpha \log{\left(\frac{M}{M_\odot}\right)} + \beta,
\label{eq:fits}
\end{equation}
for $M=M_\ast$ and $M=M_{\textrm{bulge}}$ at a range of redshifts. We only include galaxies with $M>10^{9.5} M_\odot$ in our fits, so that they are not biased by the large number of low-mass galaxies. We plot these relations in Figure \ref{Magorrian_evolution}, and give the parameters $\alpha$ and $\beta$, alongside the standard deviation of the residuals, $\epsilon$, and number of galaxies in each fit, N, in Table \ref{tab:fits}.
Both relations have a slope and normalization that increase with redshift from $z=0$ to 2, with much weaker evolution for $z>2$. Relative to the scatter in the relations, we see minimal evolution in both the black hole--bulge and black hole--total stellar mass relations from $z=0$ to 6.
This lack of evolution in the black hole--host mass relations is consistent with the findings of cosmological hydrodynamical simulations such as Horizon-AGN \citep{Volonteri2016} and {\sc BlueTides} \citep{Huang2018}.

\begin{table*}
\begin{center}
\caption{The fitting coefficients $\alpha$ and  $\beta$ (slope and normalization) of Equation \ref{eq:fits}, with the standard deviation of residuals of the scaling relations $\epsilon$ at each redshift, for $M=M_\ast$ and $M=M_{\textrm{bulge}}$. Also included is the number of galaxies that are used in each fit, N. Errors on $\alpha$ and $\beta$ are obtained from the standard deviation of 10000 bootstrap realizations. For $z\geq2$, fits are from the \textit{Tiamat} simulation, while the fits at $z<2$ use \textit{Tiamat-125-HR}. }
\label{tab:fits}
\begin{threeparttable}
\begin{tabular}{|cr|ccc|ccc|}
\hline
  &        &    \multicolumn{3}{c}{$M_\ast$} &   \multicolumn{3}{c}{$M_{\textrm{bulge}}$}\\
 $z$ &        N~~ &      $\alpha$   &     $\beta$   &  $\epsilon$ &  $\alpha$   &     $\beta$   &  $\epsilon$ \\
\hline
0 &  58997 &  1.809 $\pm$  0.002 & -11.91 $\pm$   0.02 &  0.36 &  1.624 $\pm$   0.003 & -9.72 $\pm$     0.03 &  0.32 \\
1 &  61289 &  1.833 $\pm$  0.002 & -11.90 $\pm$   0.03 &  0.28 &  1.563 $\pm$   0.002 & -8.86 $\pm$     0.03 &  0.23 \\
2 &   7503 &  1.485 $\pm$  0.005 &  -7.98 $\pm$   0.05 &  0.19 &  1.380 $\pm$    0.005 & -6.77 $\pm$     0.05 &  0.15 \\
3 &   4638 &  1.432 $\pm$  0.005 &  -7.33 $\pm$   0.05 &  0.15 &  1.377 $\pm$    0.005 & -6.69 $\pm$    0.05 &  0.13 \\
4 &   2423 &  1.392 $\pm$  0.007 &  -6.88 $\pm$   0.07 &  0.15 &  1.359 $\pm$    0.007 & -6.48 $\pm$     0.07 &  0.13 \\
5 &    917 &  1.351 $\pm$  0.016 &  -6.46 $\pm$   0.16 &  0.18 &  1.308 $\pm$   0.018 & -5.96 $\pm$     0.17 &  0.17 \\
6 &    317 &  1.367 $\pm$  0.030 &  -6.58 $\pm$   0.30 &  0.16 &  1.339 $\pm$   0.027 & -6.26 $\pm$    0.27 &  0.14 \\

\hline
\end{tabular}
\end{threeparttable}
\end{center}
\end{table*}

We find that our black hole--total stellar mass relation has similar but slightly larger scatter than the black hole--bulge relation, with the scatter in both decreasing with increasing redshift. While the black hole mass has a slightly stronger relationship with the bulge stellar mass, the black hole and total stellar mass are still tightly correlated.
The scatter in the relations is slightly larger than the 0.28 dex observed by \citet{Kormendy2013} locally. However, they are very consistent with those from the BlueTides simulation at high redshift \citep[$\simeq0.15$ dex and $\simeq0.14$ dex for the black hole--bulge and total stellar mass relations, respectively;][]{Huang2018}.
The scatter decreases with increasing stellar mass---including only galaxies with $M_{\ast/\textrm{bulge}}/M_\odot>(10^{10},10^{10.5})$ reduces the scatter to $\epsilon=(0.30,0.20)$ dex and $\epsilon=(0.23, 0.16)$ dex, for the $z=0$ black hole--total stellar mass and bulge mass relations, respectively.

\begin{figure}
\begin{center}
\includegraphics[scale=0.9]{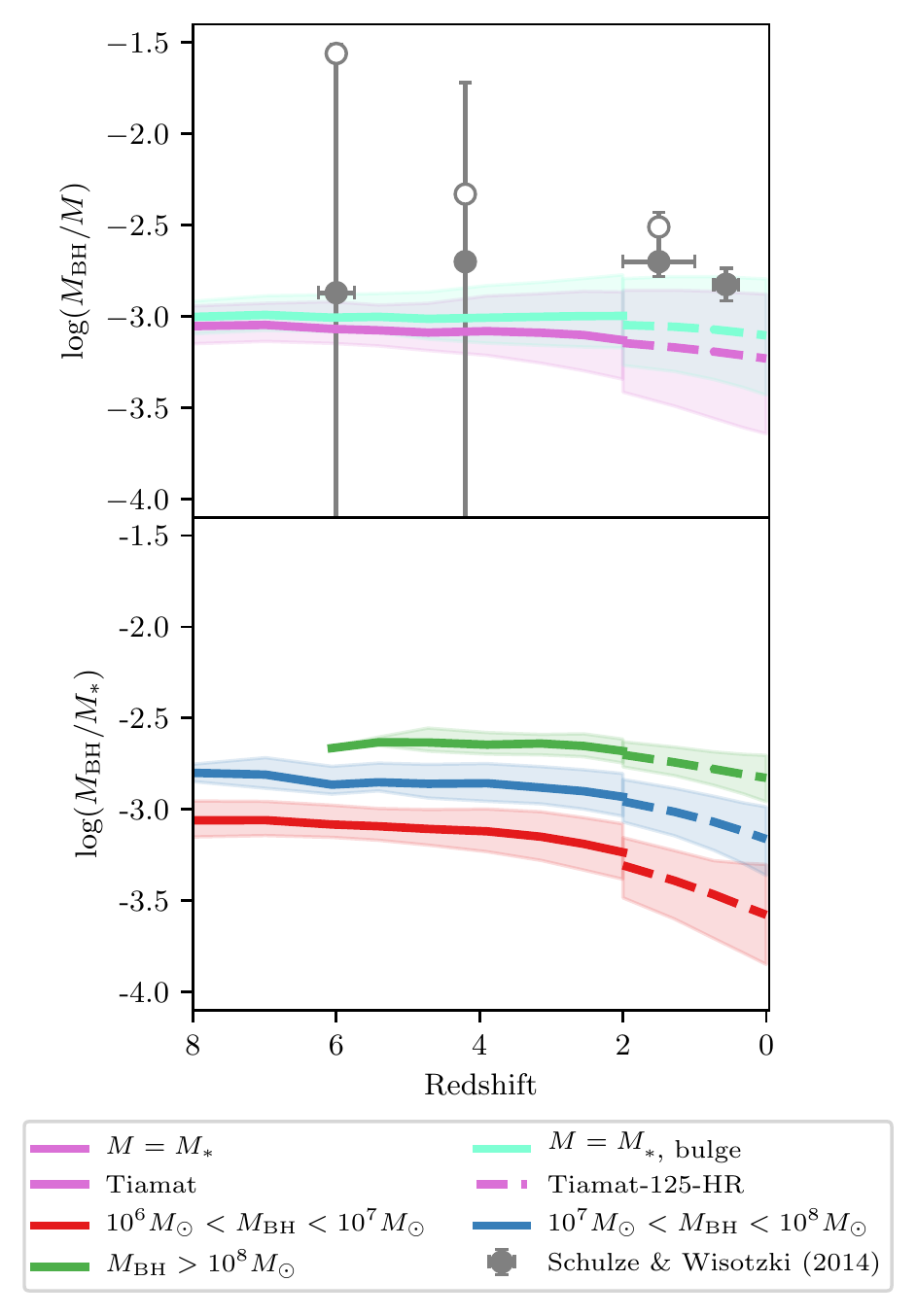}
\caption{\textit{Upper panel:} the ratio of black hole to total stellar mass and bulge stellar mass as a function of redshift, for galaxies with $M_{\textrm{BH}}>10^6M_\odot$. Results from the observational analysis of \citet{Schulze2014} are also plotted, showing the apparent relation (open grey points), and the intrinsic relation, where selection effects have been accounted for (closed grey points). \textit{Lower panel:} the ratio of black hole to total stellar mass as a function of redshift for various black hole masses; note that these black hole mass cuts are made at each redshift, and so this is not showing the evolution of ratio over time for black holes of a given size. The median ratio is shown with the solid line, and the region between the 16th and 84th percentile range shaded. }
\label{MeanBHBulge}
\end{center}
\end{figure}

Figure \ref{MeanBHBulge} shows the median $M_{\mathrm{BH}}/M$ as a function of redshift for galaxies with $M_{\textrm{BH}}>10^6M_\odot$, for $M=M_\ast$ and $M=M_{\mathrm{bulge}}$. The figure shows no statistically-significant evolution in the median $M_{\mathrm{BH}}/M_{\mathrm{bulge}}$ and $M_{\mathrm{BH}}/M_\ast$ out to $z\simeq8$. This is consistent with current high redshift observations; when selection effects are accounted for, the observations at high redshift are consistent with no cosmological evolution in these relations \citep{Schulze2014}.

Our model predicts no significant evolution in the black hole--host mass relations, with the scatter in the relations decreasing at the highest redshifts. This indicates that there is a connection between the growth of black holes and their host galaxies. Indeed, our model includes joint triggering of star formation and black hole growth during galaxy mergers, and black hole feedback which regulates star formation, meaning that the co-evolution of black holes and galaxies is implicit in our model.
This is not consistent with the scenario proposed by \citet{Peng2007} and \citet{Jahnke2011}, for example, where
the black hole and galaxy growth is uncorrelated and the relationships are generated naturally within a merger driven galaxy evolution framework, due to a central-limit-like tendency. 

Figure \ref{MeanBHBulge} also shows the median $M_{\mathrm{BH}}/M_\ast$ ratio as a function of redshift with galaxies split into black hole mass bins. This shows that lower mass black holes have lower $M_{\mathrm{BH}}/M_\ast$ ratios than higher mass black holes. For example, at high redshifts ($z>2$), the median $M_{\mathrm{BH}}/M_\ast$ ratio for black holes with $10^7<M_{\mathrm{BH}}/M_\odot<10^8$ is higher than those with $10^6<M_{\mathrm{BH}}/M_\odot<10^7$ by $\sim0.25$ dex, with that for black holes with $M_{\mathrm{BH}}/M_\odot>10^8$ being a further $\sim0.25$ dex higher. This will lead to a notable selection bias, since when observing the most massive black holes, the measured $M_{\mathrm{BH}}/M_\ast$ ratio will be higher than that of the entire population. This is generally expected for any sample selected by black hole mass or luminosity where the scatter in the relation is large \citep[e.g.][]{Lauer2007}.

\begin{figure}
\begin{center}
\includegraphics[scale=0.9]{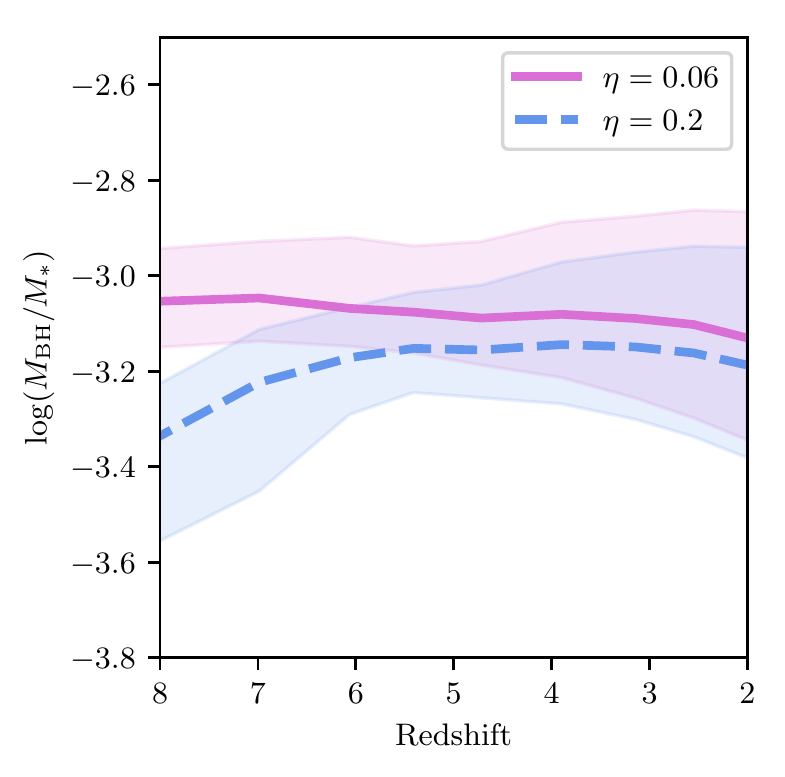}
\caption{The ratio of black hole to total stellar mass as a function of redshift for galaxies with $M_{\textrm{BH}}>10^6M_\odot$, for the best model ($\eta=0.06$) and an otherwise identical model with $\eta=0.2$. The median ratio is shown with the solid line, and the region between the 16th and 84th percentile range shaded. }
\label{MeanBHBulge_eta}
\end{center}
\end{figure}

Finally, we note an interesting effect of changing the parameter controlling the black hole efficiency for converting mass to energy, $\eta$ (see Section \ref{sec:BHGrowth}). The median black hole--stellar mass ratio for our best model is shown in Figure \ref{MeanBHBulge_eta}, alongside \meraxes run with $\eta=0.2$ instead of 0.06, with all other parameters unchanged. For $\eta=0.2$, the median black hole--stellar mass ratio decreases at redshifts $z\gtrsim6$, instead of remaining constant with redshift as in the  $\eta=0.06$ model. This effect is not seen by adjusting any of the other black hole parameters we tune in the model (Table \ref{tab:Parameters}). We investigate the cause of this high-redshift decrease in the black hole--host relation by considering the Eddington limit:
\begin{equation}
M_{\textrm{Edd}}=M_{\textrm{BH}} \left( \exp \left( \frac{\Delta t}{\eta t_{\textrm{Edd}}}\right) -1 \right),
\end{equation}
the maximal mass by which a black hole with mass $M_{\textrm{BH}}$ can grow in the model between snapshots of width $\Delta t$ (see Equation \ref{eq:Edd}).  Increasing $\eta$ from 0.06 to 0.2 decreases the Eddington limit. This results in many black holes having Eddington-limited growth at the highest redshifts ($z\gtrsim6$), which is not the case for the $\eta=0.06$ model.
This causes black holes to grow slower than their host galaxies at high redshifts, resulting in a decreased black hole--stellar mass ratio. Observing the high-redshift black hole--stellar mass relation may therefore probe the Eddington limit and the efficiency of  black holes in converting mass to energy.

\subsection{Black hole growth mechanisms}
\label{sec:BHGrowthModes}
\begin{figure*}
\begin{center}
\includegraphics[scale=0.9]{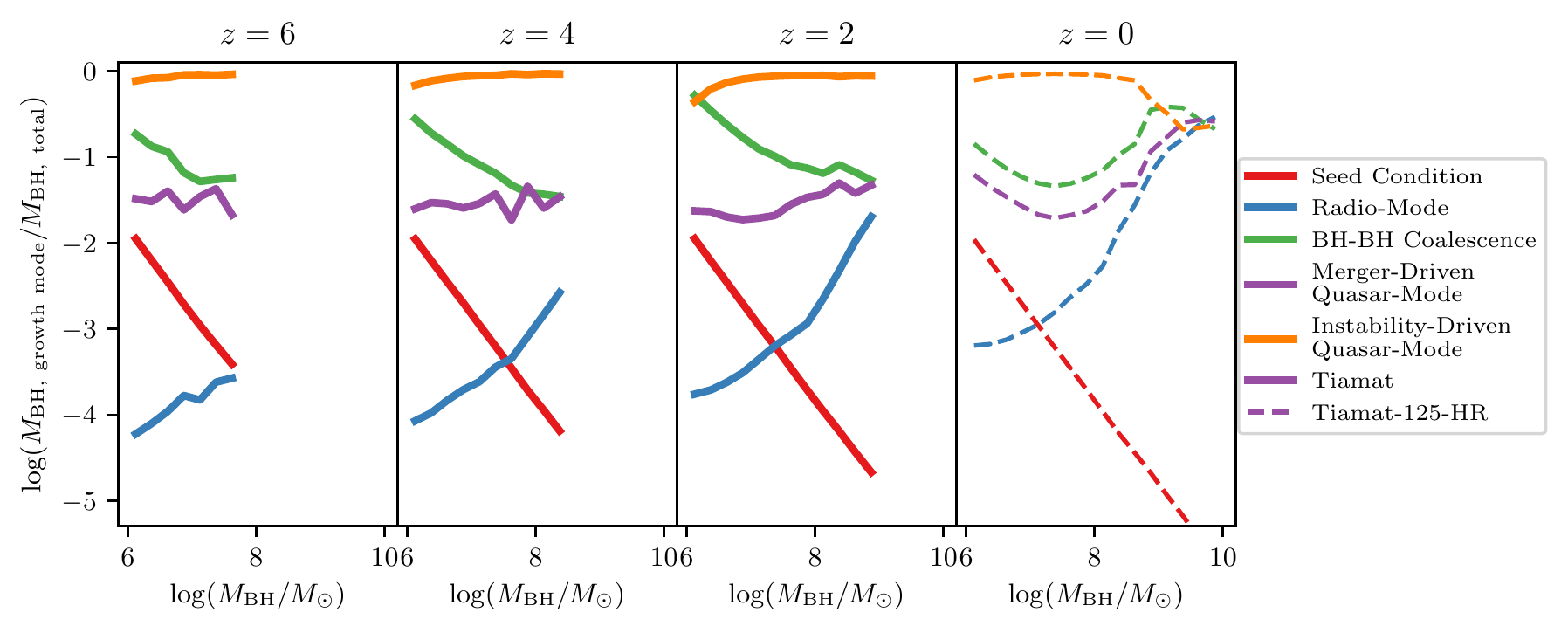}
\caption{The average fraction of black hole mass formed through each of the growth mechanisms in \meraxes relative to the total black hole mass, in black hole mass bins of 0.25 dex, for a range of redshifts. Note that these are cumulative fractions, and not the fraction of growth produced by each mechanism \textit{at} that redshift. The model with merger-driven black hole growth efficiency $k_c=k_i=0.005$ is shown.}
\label{BHGrowthModes}
\end{center}
\end{figure*}
We consider the cumulative fraction of black hole mass formed through each of the mechanisms in our model: black hole seeding, merger-driven quasar-mode accretion, instability-driven quasar-mode accretion, radio-mode accretion and black hole--black hole coalescence in galaxy mergers. 
We plot these as a function of black hole mass at a range of redshifts in Figure \ref{BHGrowthModes}. 
On average, instabilities grow the majority of mass in black holes at all redshifts, except for galaxies with $M_{\textrm{BH}}>10^9M_\odot$ at $z\simeq0$, whose black hole growth becomes dominated by galaxy mergers.
Radio-mode growth slowly increases in significance with redshift, yet still has only contributed to a small proportion of the total black hole mass by $z=0$, except at the highest masses; this is discussed in \citetalias{Qin2017}. 
Note that we consider growth from disc instabilities that are triggered by earlier galaxy mergers as growth via the instability-driven mode, and do not treat them in a more detailed manner as in \citet{IzquierdoVillalba2019}, for example.

\begin{figure}
\begin{center}
\includegraphics[scale=0.9]{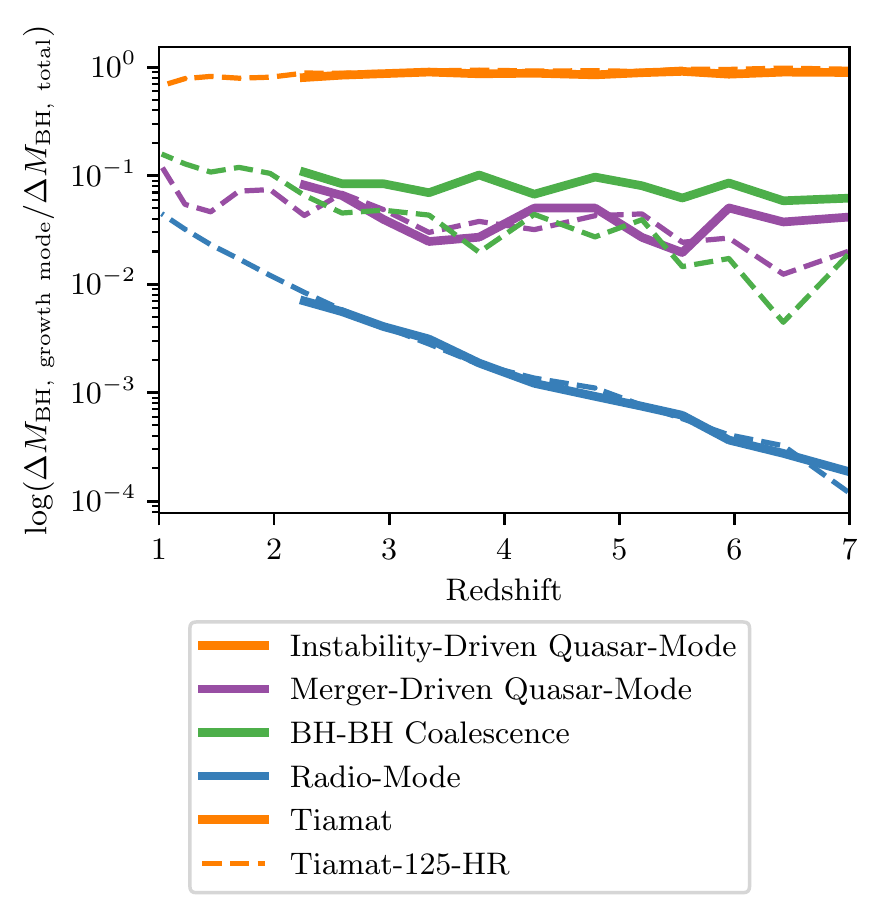}
\caption{The instantaneous fraction of black hole mass formed through each of the growth mechanisms in \meraxes relative to the total black hole mass, as a function of redshift from $z=7$ to 1. We take the `instantaneous' fraction to be the fraction of growth between the specified redshift and the simulation snapshot immediately preceding it. We do not include the seed mechanism, as that `growth' occurs only once for each black hole. The model with merger-driven black hole growth efficiency $k_c=k_i=0.005$ is shown. We do not show the results at $z<1$, as the model produces unreliable black hole accretion rates at such low redshifts (see Figure \ref{BHARD} and its discussion).}
\label{BHGrowthModes_redshift}
\end{center}
\end{figure}

We also consider the instantaneous growth fractions of black hole mass formed through each mechanism as a function of redshift, as shown in Figure \ref{BHGrowthModes_redshift}. Here we take the `instantaneous' fraction to be the fraction of growth caused by a mechanism between the specified redshift and the simulation snapshot immediately preceding it. 
As discussed in Section \ref{sec:QLFs}, the model produces unreliable black hole accretion rates at $z<1$, and so we only consider these black hole growth rates at $z>1$.
Figure \ref{BHGrowthModes_redshift} shows that the instability-driven growth mode is the dominant growth mechanism, on average, at all redshifts, regardless of black hole mass. The merger-driven quasar mode and black hole--black hole coalescence mode are sub-dominant at all redshifts. The radio-mode grows more mass at low redshift and in the most massive galaxies, with the percentage of total instantaneous black hole growth from this mode increasing from only 0.1 per cent at $z=5$ to almost 5 per cent at $z\simeq1$. 

Our finding that mergers are not the dominant mechanism for growing black holes is in agreement with a range of observations.  
For example, \citet{Koss2010} find that only 25 per cent of local ($z<0.05$), moderate luminosity X-ray AGN show signs of mergers, though the fraction is much higher for luminous AGN \citep{Hong2015}. From $z\simeq0.3$-1.0, \citet{Cisternas2010} find that the vast majority ($>85$ per cent) of X-ray selected AGN do not show signs of mergers, suggesting that the bulk of their black hole accretion has been triggered by some other mechanism. This is also consistent with the findings of \citet{Georgakakis2009} who claim that a large fraction of AGN at $z\simeq1$ are triggered by processes other than major mergers, as do \citet{Villforth2018} at $z\simeq0.9$, and \citet{Schawinski2012}, \citet{Mechtley2016}, \citet{DelMoro2015} and \citet{Marian2019} for AGN at $z\simeq2$.

Our result that disc instabilities cause the majority of black hole growth is also consistent with predictions from other simulations.
In the GALFORM semi-analytic model, \citet{Fanidakis2011} found that the growth of black holes is dominated by accretion due to disc instabilities, with the fraction of mass in black holes produced by disc instabilities more than an order of magnitude larger than that produced by mergers, at all redshifts.
Using an updated GALFORM model, \citet{Griffin2019} found that accretion of hot gas dominates the growth of black holes at $z<2$, with disc-instabilities dominant at higher redshifts.
\citet{Hirschmann2012} found that instability-driven black hole growth was required to reproduce AGN downsizing, and that while major mergers are the dominant trigger for luminous AGN, especially at high redshift, disc instabilities cause the majority of black hole growth in moderately luminous Seyfert galaxies at low redshift.
\citet{Menci2014} find that in their semi-analytic model disc instabilities can provide enough black hole accretion to reproduce the observed AGN luminosity functions up to $z\approx4.5$, but are not likely to be dominant for the highest luminosity AGN or at the highest redshifts. In contrast, \citet{Shirakata2018} find that the primary trigger of AGN at $z\leq4$ in their semi-analytic model is mergers, while disc instabilities are essential for fuelling moderate luminosity AGN at higher redshifts.
The hydrodynamical simulation Horizon-AGN found that only $\sim35$ per cent of black hole mass in local massive galaxies is directly attributable to merging, with the majority of black hole growth instead growing via secular processes \citep{Martin2018}.
The Magneticum Pathfinder Simulation also found that merger events are not the dominant fuelling mechanism for black holes in $z=0$--2, with merger fractions less than 20 per cent, except for very luminous quasars at $z\simeq2$ \citep{Steinborn2018}.

\begin{figure*}
\begin{center}
\includegraphics[scale=0.9]{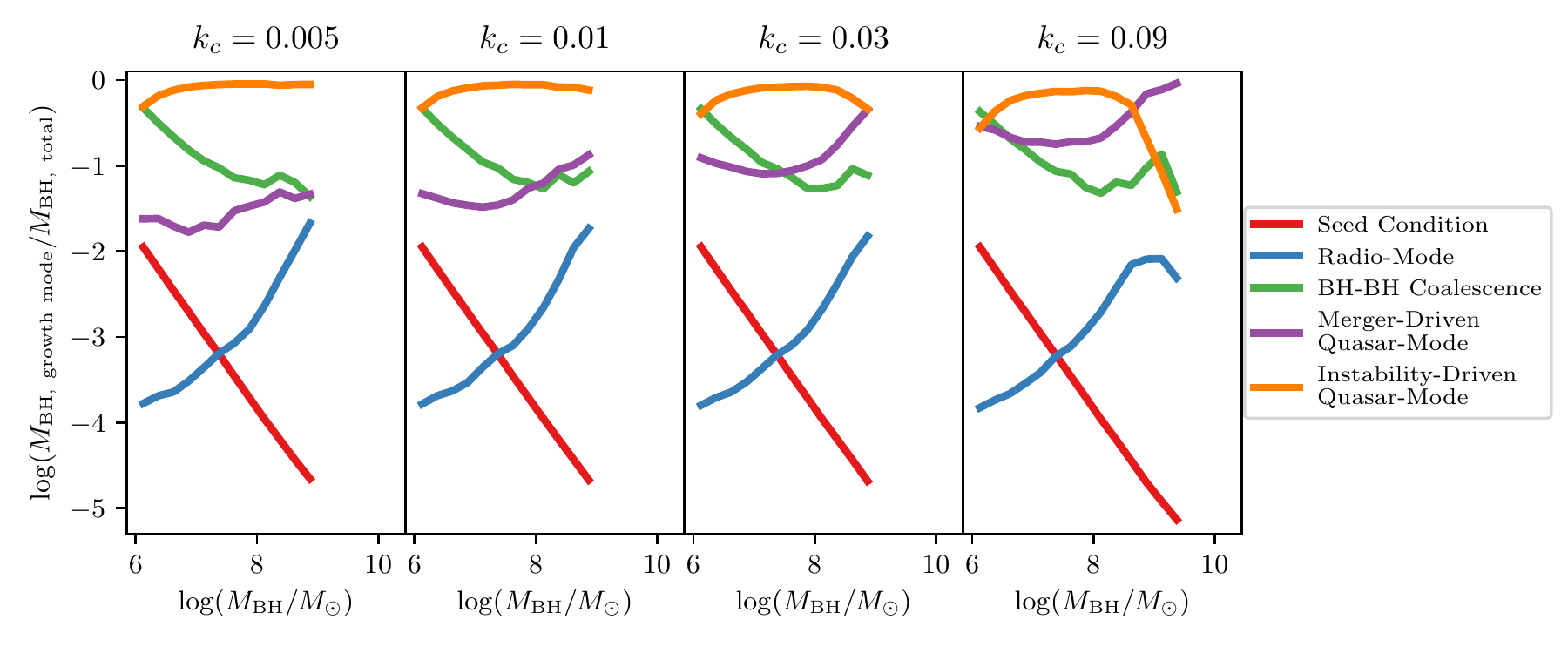}
\caption{The average fraction of black hole mass formed through each of the growth mechanisms in \meraxes relative to the total black hole mass by $z=2$, in black hole mass bins of 0.25 dex, for different merger-driven black hole growth efficiencies: $k_c=0.005$, 0.01, 0.03 and 0.09. These parameters were all found during the model tuning to reproduce the observations well, however the larger values of $k_c$ produce a black hole growth history that is larger than observed. Increasing $k_c$ increases the contribution of the merger-driven mode to growing black holes, but the instability-driven mode is still dominant except for at the lowest and highest black hole masses. Note that these are cumulative fractions, and not the fraction of growth produced by each mechanism \textit{at} $z=2$.}
\label{BHGrowthModes_modelComparison}
\end{center}
\end{figure*}

Finally, we comment on the effect of the efficiency parameters for merger-driven and instability-driven black hole growth in the model, $k_c$ and $k_i$ respectively (see Equation \ref{eq:BHgrowth}). We find $k_i=0.005$ from tuning the model, whereas $k_c$ is less constrained, with $k_c=0.005$, 0.01, 0.03 and 0.09 producing reasonable model results (though the larger values of $k_c$ produce a black hole growth history that is too large; see Section \ref{Verification} and Figure \ref{BHARD}). Having a merger growth efficiency that is twice, six times or even 18 times larger than the instability-driven growth efficiency may have an effect on the conclusions outlined above, which use the model $k_c=k_i=0.005$. We therefore plot the cumulative fraction of black hole mass formed through each of the mechanisms at $z=2$ for all four models, $k_c=0.005$, 0.01, 0.03 and 0.09 (Figure \ref{BHGrowthModes_modelComparison}). We find, as expected, that models with larger $k_c$ result in more merger-driven growth. For $k_c=0.01$, the instability-driven mode still dominates at $z=2$, while for $k_c=0.03$, the merger-driven mode begins to dominate at the highest black hole masses, $M_{\textrm{BH}}\simeq10^9M_\odot$. For the model with $k_c=0.09$, the merger-driven mode contributes even more black hole growth, but is still not the dominant growth mode for $10^6<M_{\textrm{BH}}/M_\odot<10^9$ black holes. 
The  \textit{Tiamat-125-HR} simulation at $z=0$ shows the same trend, with the merger-driven growth mode becoming more dominant as $k_c$ increases; for the most extreme case of $k_c=0.09$, the merger-driven mode remains the dominant growth mode for $M_{\textrm{BH}}>10^9M_\odot$ black holes, however the instability-driven mode is still the main source of growth for smaller black holes.
Thus, while the efficiency parameter for merger-driven growth has some effect on the relative distributions of the instability-driven and merger-driven growth modes, the instability-driven mode is still dominant for the majority of black holes, even if the merger growth efficiency is as much as 18 times larger than the secular growth efficiency.  

\subsection{The morphology dependence of the black hole--host mass relations}
\label{sec:Morphology}

\begin{figure*}
\begin{center}
\includegraphics[scale=0.9]{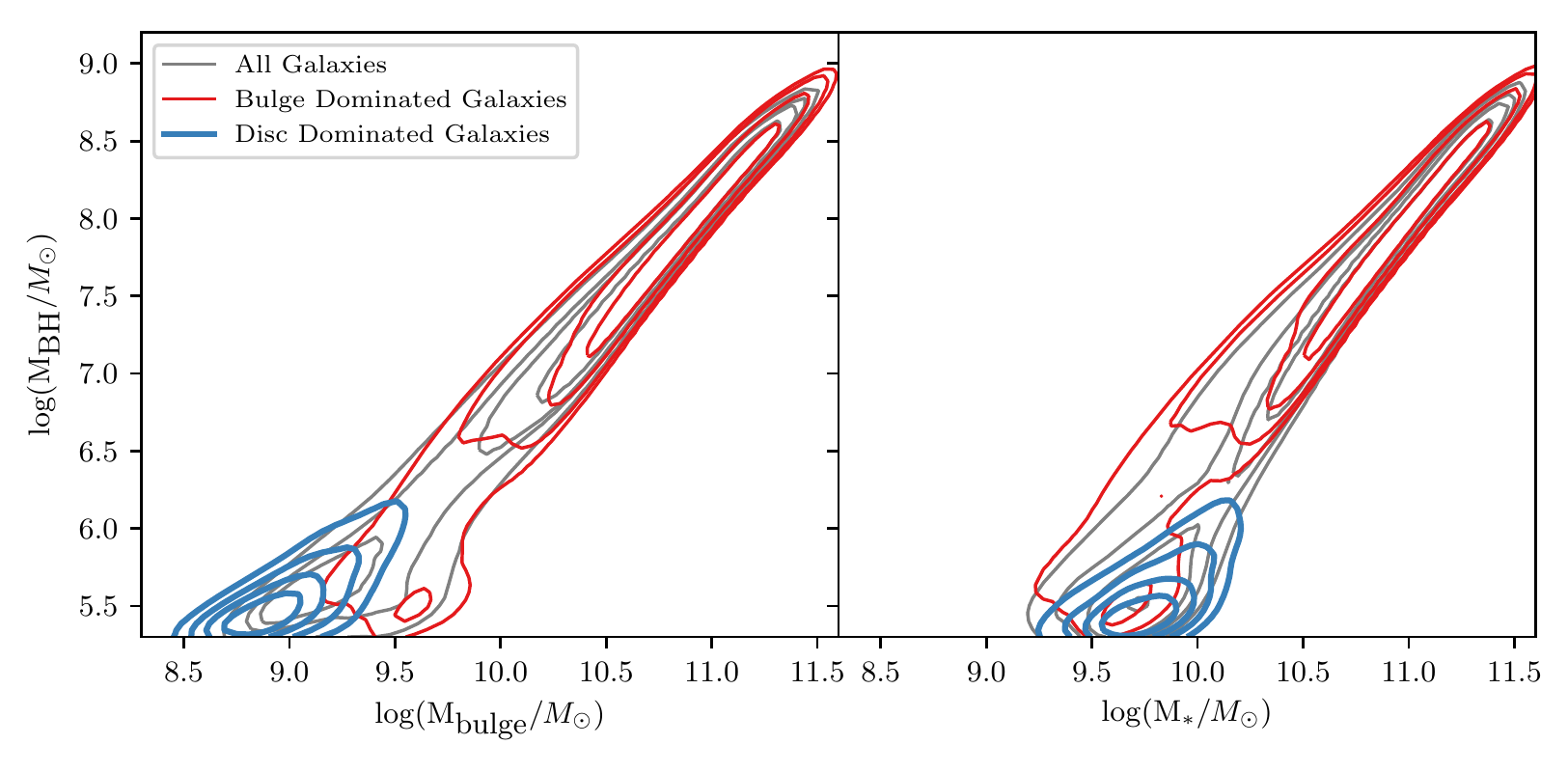}
\caption{\textit{Left panel:} the black hole--bulge mass relation, and \textit{Right panel:} the black hole--total stellar mass relation, for \meraxes galaxies at $z=0$. Galaxies are split into bulge-dominated galaxies ($B/T>0.7$; red contours) and disc-dominated galaxies ($B/T<0.3$; blue contours), with the distribution for all galaxies also shown (grey contours). Contours show regions containing probability distributions of 20, 40, 60 and 80 per cent.}
\label{Morphology}
\end{center}
\end{figure*}

A popular explanation for the black hole--host correlations is that major mergers drive the growth of both black holes and bulges \citep[e.g.][]{Haehnelt2000,Croton2006b}. If this were the case, one would expect that black holes would only correlate with galaxy properties directly related to the merger process, such as bulge mass, and not, for example, total stellar mass.  \citet{Simmons2017} consider a sample of 101 disc-dominated AGN hosts from the SDSS, which they assume must have a major merger-free history since $z\simeq2$. They found that these galaxies lie on the typical $M_{\textrm{BH}}$--$M_\ast$ relation, but lie offset to the left of the $M_{\textrm{BH}}$--$M_{\textrm{bulge}}$ relation. This indicates that the substantial and ongoing black hole growth in these merger-free disc galaxies must be due to a process other than major mergers, and that major mergers cannot be the primary mechanism behind the black hole--host correlations.

We plot the $M_{\textrm{BH}}$--$M_\ast$ and $M_{\textrm{BH}}$--$M_{\textrm{bulge}}$ relation for disc-dominated and bulge-dominated galaxies at $z=0$ in Figure \ref{Morphology}.
Our simulated disc galaxies lie on the $M_{\textrm{BH}}$--$M_\ast$ relation, but lie offset to the left of the $M_{\textrm{BH}}$--$M_{\textrm{bulge}}$ relation, as they have small bulges relative to their black hole mass. This is consistent with the \citet{Simmons2017} observations, and the results from the Horizon-AGN hydrodynamical simulation \citep{Martin2018}. However, we see a less significant offset, which occurs at lower black hole masses than \citet{Simmons2017} and \citet{Martin2018}, since the black holes in our disc-dominated galaxies are less massive in comparison.
\citet{Mutlu-Pakdil2017} also find no dependence of the $M_{\textrm{BH}}$--$M_\ast$ relation on galaxy type in the Illustris hydrodynamical simulation.
\citet{Martin2018} suggest that major mergers therefore cannot be primarily responsible for feeding black holes, otherwise major-merger free disc galaxies should have less massive black holes than are observed and simulated. 
This is consistent with our finding that the instability-driven mode is the dominant growth mechanism for black holes (see Section \ref{sec:BHGrowthModes}).

\section{Conclusions}
\label{Conclusion}
We use the \meraxes semi-analytic model to investigate the evolution of black holes and their relations to their host galaxies. We find the following key predictions of our model:
\begin{itemize}
    \item There is minimal statistically-significant evolution in the black hole--bulge and black hole--total stellar mass relations out to high redshifts ($z\simeq8$).
    \item The black hole--total stellar mass relation has similar but slightly larger scatter than the black hole--bulge relation, with the scatter in both decreasing with increasing redshift. This indicates that the growth of galaxies, bulges and black holes are all tightly related, even at the highest redshifts.
    \item Higher mass black holes have higher black hole--total stellar mass ratios, leading to a significant selection effect in measurements of this ratio when observing only the most massive black holes.
    \item The instability-driven or secular quasar-mode growth is the dominant growth mechanism for black holes at all redshifts.  The contribution from merger-driven quasar-mode growth only becomes significant at low redshift for black holes with $M_{\textrm{BH}} \gtrsim 10^{9}M_\odot$.
    \item Disc-dominated galaxies lie on the black hole--total stellar mass relation, but lie offset from the black hole--bulge mass relation.
\end{itemize}

Our simulation is limited in making predictions for the highest redshift quasars at $z=6$--7 due to the simulation box size and resolution. In future work we will run \meraxes on larger N-body simulations in order to make predictions for these objects.

\section*{Acknowledgements}
We thank the anonymous referee for their constructive comments.
This research was supported by the Australian Research Council Centre of Excellence for All Sky Astrophysics in 3 Dimensions (ASTRO 3D), through project number CE170100013.
This work was performed on the OzSTAR national facility at Swinburne University of Technology. OzSTAR is funded by Swinburne University of Technology and the National Collaborative Research Infrastructure Strategy (NCRIS).
MAM acknowledges the support of an Australian Government Research Training Program (RTP) Scholarship.




\bibliography{Paper_BH_Arxiv.bib} 



\appendix
\section{Calibration}
\label{sec:Appendix}
\begin{figure*}
\begin{center}
\includegraphics[scale=0.9]{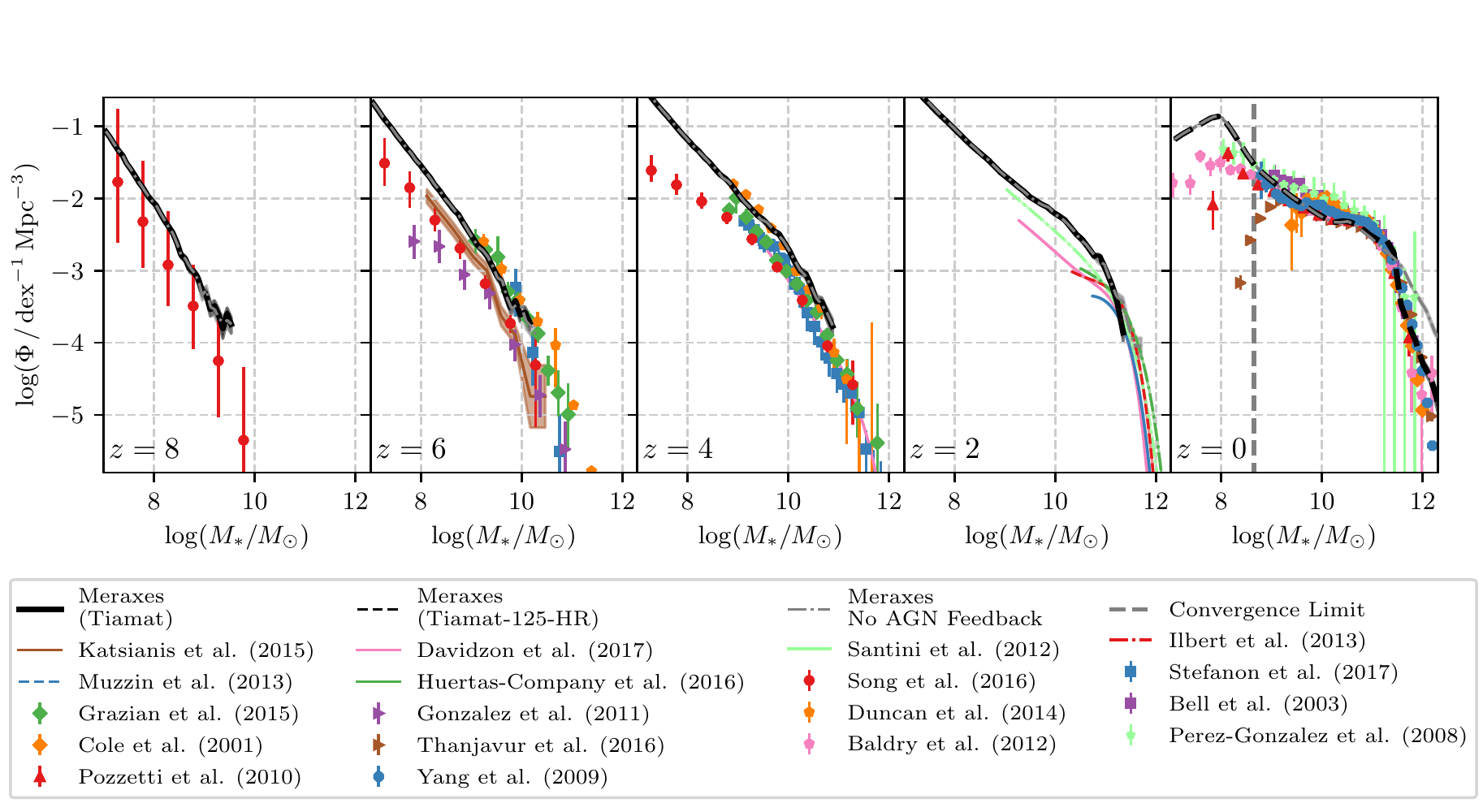}
\caption{Galaxy stellar mass functions at $0<z<8$ from our best \meraxes model (black) applied to \textit{Tiamat} (solid) and \textit{Tiamat-125-HR} (dashed), compared to a range of observational data (see legend). \meraxes is calibrated such that these observed stellar mass functions are reproduced. The vertical grey dotted line indicates the stellar mass below which  \textit{Tiamat} and \textit{Tiamat-125-HR} are not converged, and thus where galaxies from \textit{Tiamat-125-HR} can be subject to resolution effects \citepalias[see][]{Marshall2019}.
Also shown are the stellar mass functions produced by \meraxes when AGN feedback is switched off (grey dot-dashed).
}
\label{SMF}
\end{center}
\end{figure*}

\begin{figure*}
\begin{center}
\includegraphics[scale=0.9]{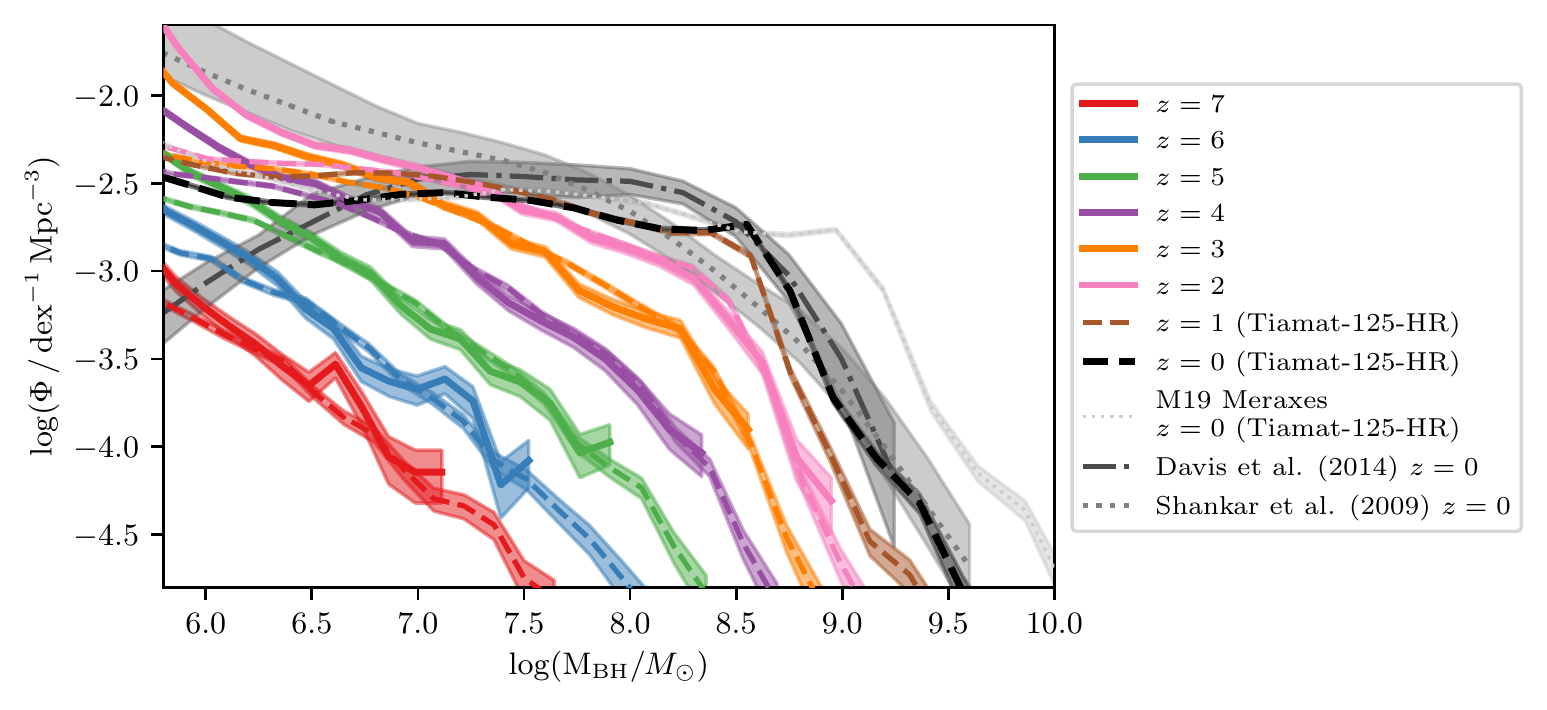}
\caption{Black hole mass functions at $0<z<7$ from our best \meraxes model applied to \textit{Tiamat} (solid) and \textit{Tiamat-125-HR} (dashed). \meraxes is calibrated to best reproduce the \citet{Shankar2009} and \citet{Davis2014} observed black hole mass functions at $z=0$, which are also shown (see legend). We also plot the $z=0$ black hole mass function from the \citetalias{Marshall2019} \meraxes model, showing that this overpredicted the observed black hole mass functions.}
\label{BHMF}
\end{center}
\end{figure*}

We calibrate the free parameters in \meraxes to match the observed stellar mass functions at $z=0$--8 and the \citet{Shankar2009} and \citet{Davis2014} black hole mass function at $z=0$, shown here in Figures \ref{SMF} and \ref{BHMF}. 
The black hole mass functions produced by \textit{Tiamat} and \textit{Tiamat-125-HR} are converged at $z=2$ for black holes with mass $M_{\textrm{BH}}>10^{7.1}M_\odot$ \citepalias[see Figure \ref{BHMF} and][]{Marshall2019}, with \textit{Tiamat-125-HR} producing more low-mass black holes. We therefore focus on matching the observed black hole mass functions at $M_{\textrm{BH}}>10^{7.1}M_\odot$. While the \citet{Shankar2009} and \citet{Davis2014} relations are different, particularly at $M_{\textrm{BH}}\sim10^{8.5}M_\odot$, they are similar relative to the freedom we have in adjusting our model black hole mass function, and so when calibrating we found the most reasonable fit to both.

In Figure \ref{SMF} we also plot the \meraxes stellar mass function produced when AGN feedback is switched off. This shows that AGN feedback has no effect on galaxies in \textit{Tiamat} at $z\geq2$, but suppresses the growth of the most massive galaxies at lower redshifts as seen in \textit{Tiamat-125-HR}.

\section{Consistency of Tiamat and Tiamat-125-HR}
\label{sec:Appendix2}
Throughout this work, we use the higher resolution \textit{Tiamat} simulation at $z\geq2$, and \textit{Tiamat-125-HR} for $z<2$, where \textit{Tiamat} is unavailable.
We find that the results discussed in this paper are generally consistent between the two simulations at $z\simeq2$, and so in general we are confident that any redshift evolution we find at $z<2$ is not caused by a change in simulation.

\begin{figure*}
\begin{center}
\includegraphics[scale=0.9]{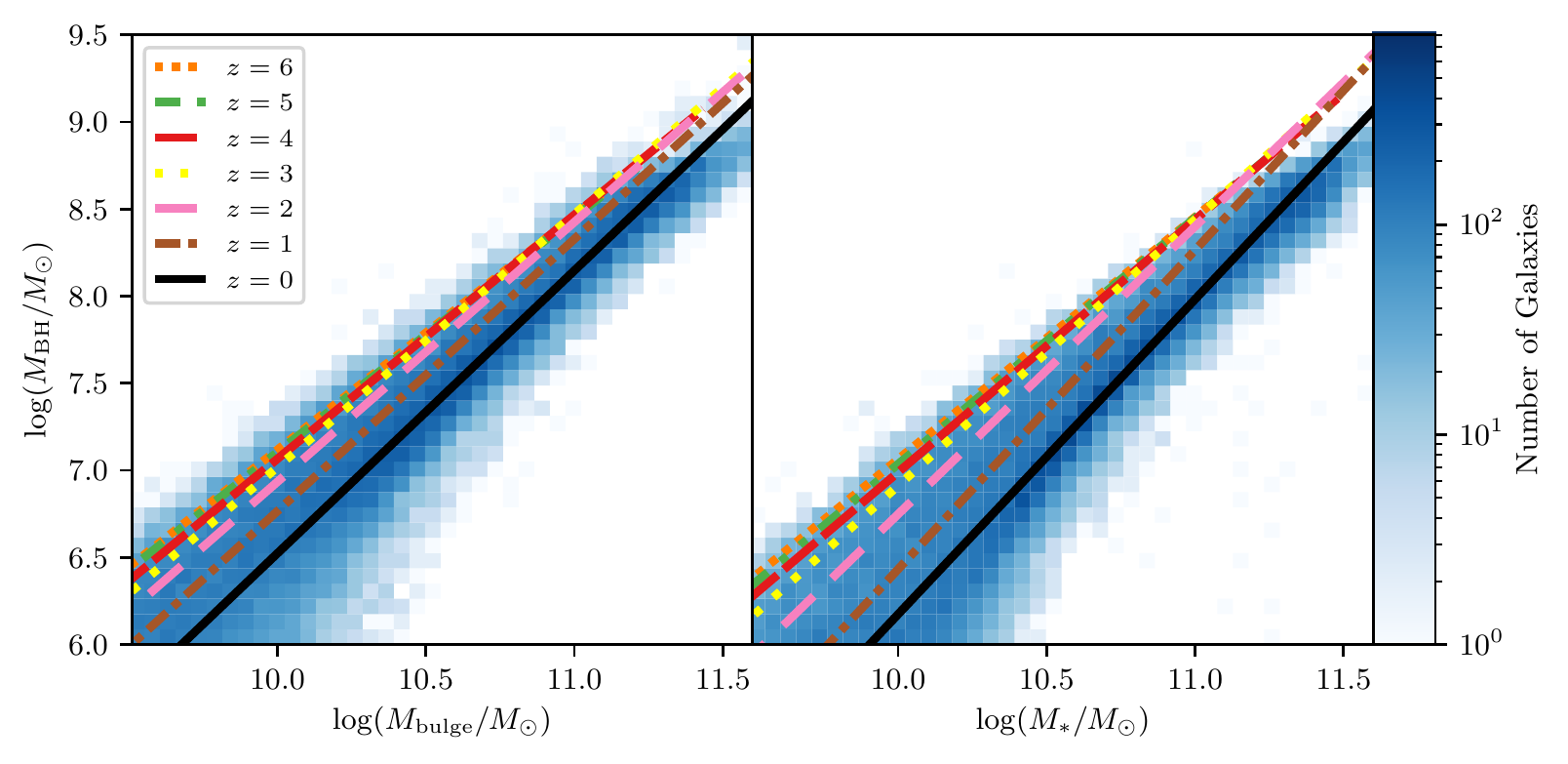}
\caption{Lines of best-fit to the black hole--bulge mass \textit{(left panel)} and black hole--total stellar mass \textit{(right panel)} relations at a range of redshifts, using the \textit{Tiamat-125-HR} simulation. The blue density plot shows the $z=0$ distribution. This shows similar results to those found using the \textit{Tiamat} simulation at $z\geq2$ (Figure \ref{Magorrian_evolution}), with a slightly milder evolution from $z=2$ to $z=1$.
}
\label{Magorrian_evolution_T125}
\end{center}
\end{figure*}

However, one notable result is that the best-fitting black hole--stellar mass relations of Figure \ref{Magorrian_evolution} change rapidly between $z=2$ (using \textit{Tiamat}) and $z=1$ (using \textit{Tiamat-125-HR}).  To verify that this jump is not purely a result of the simulation change, we show the best-fitting relations from $z=6$--0 using \textit{Tiamat-125-HR} (Figure \ref{Magorrian_evolution_T125}).
The \textit{Tiamat-125-HR} simulation shows similar results to those found using \textit{Tiamat} at $z\geq2$ (Figure \ref{Magorrian_evolution}), with a slightly milder but still relatively rapid evolution from $z=2$ to $z=1$. The qualitative result of the evolution being insignificant relative to the scatter in the relation still holds. Thus, while the change in simulation slightly amplifies the rapid change in the black hole--stellar mass relations from $z=2$ to $z=1$, this does not change our conclusions. We also note that where the black hole mass functions are converged ($M_{\textrm{BH}}>10^{7.1}M_\odot$), the black hole--stellar mass relations are in good agreement between the two simulations.

\bsp	
\label{lastpage}
\end{document}